{}
\newcommand{\ACCELERATOR}{DSTAR}

\documentclass[sigconf]{acmart}

\renewcommand\footnotetextcopyrightpermission[1]{}

\usepackage{amsmath,amsfonts}
\usepackage{algorithmic}
\usepackage{graphicx}
\usepackage{textcomp}
\usepackage{xcolor}
\usepackage{multirow}
\usepackage{listings}
\usepackage{booktabs}
\usepackage{tabularx}
\usepackage{multicol}
\usepackage{enumitem}
\usepackage{tikz}
\usepackage{amsmath}
\usepackage{caption}
\usepackage{subcaption}
\usepackage{rotating}
\usepackage{pifont}
\usepackage{adjustbox}
\usepackage{url}
\usepackage{svg}
\usepackage{makecell}
\usepackage{hyperref}
\usepackage{threeparttable}

\definecolor{titleblue}{RGB}{0, 76, 151}   
\definecolor{reviewerA}{RGB}{200, 50, 50}   
\definecolor{reviewerB}{RGB}{50, 150, 50}   
\definecolor{reviewerC}{RGB}{50, 100, 200}  
\definecolor{reviewerD}{RGB}{180, 80, 180}  
\definecolor{reviewerE}{RGB}{220, 140, 0}   
\definecolor{reviewerF}{RGB}{0, 170, 170}   
\definecolor{general}{RGB}{180, 215, 50}      

\AtBeginDocument{%
  }

\setcopyright{none} 
\copyrightyear{2026}
\acmYear{2026}
\acmDOI{}

\acmConference[MICRO 2026]{The 58th IEEE/ACM International Symposium on Microarchitecture}{October 31--November 04, 2026}{Athens, Greece}

\acmISBN{}

\begin{document}

\title{\ACCELERATOR: Accelerating Diffusion Transformers via Spatial and Temporal Redundancy Reduction}

\author{Chi Zhang}
\affiliation{
\department{School of Computer Science}
  \institution{Shanghai Jiao Tong University}
  \city{Shanghai}
  \country{China}
}
\email{zhang-chi@sjtu.edu.cn}

\author{Jieru Zhao}
\authornote{Corresponding author.}
\affiliation{
\department{School of Computer Science}
  \institution{Shanghai Jiao Tong University}
  \city{Shanghai}
  \country{China}
}
\email{zhao-jieru@sjtu.edu.cn}

\author{Yu Feng}
\affiliation{
\department{School of Computer Science}
  \institution{Shanghai Jiao Tong University}
  \city{Shanghai}
  \country{China}
}
\email{y-feng@sjtu.edu.cn}

\author{Chen Zhang}
\affiliation{
\department{School of Electronic Information and Electrical Engineering}
  \institution{Shanghai Jiao Tong University}
  \city{Shanghai}
  \country{China}
}
\email{chenzhang.sjtu@sjtu.edu.cn}

\author{Quan Chen}
\affiliation{
\department{School of Computer Science}
  \institution{Shanghai Jiao Tong University}
  \city{Shanghai}
  \country{China}
}
\email{chen-quan@cs.sjtu.edu.cn}

\author{Minyi Guo}
\affiliation{
  \institution{Guizhou Provincial Laboratory of Big Data, College of Computer Science and Technology, Guizhou University}
  \city{Guizhou}
  \country{China}
}
\affiliation{
  \institution{School of Computer Science, SJTU}
  \city{Shanghai}
  \country{China}
}
\email{guo-my@cs.sjtu.edu.cn}





\begin{abstract}

Diffusion Transformers (DiTs) have been widely used in many tasks, 
including image synthesis, video generation, and content editing.
However, their multi-iteration inference process leads to performance inefficiency and high energy consumption. Existing acceleration methods primarily focus on reducing temporal redundancy between adjacent timesteps,
but often overlook the specific features of DiTs. As a result, these approaches either
suffer from great accuracy degradation or fail to achieve high efficiency.

We present DSTAR, a software-hardware co-design framework that accelerates DiT inference by reducing spatial and temporal redundancy. At the algorithmic level, DSTAR introduces a fine-grained mixed-precision quantization method for differential activations in linear operations, significantly increasing the proportion of low-bit computations. Additionally, DSTAR incorporates a sparse attention reuse mechanism to minimize redundant computation in attention layers.
For architectural support, we design a specialized hardware accelerator which
achieves high efficiency in both latency and energy consumption. 
Evaluation on seven typical DiTs demonstrates that DSTAR achieves up to $7.33\times$ latency speedup and $41.89\times$ energy savings compared to an NVIDIA A100 GPU, and achieves up to $2.54\times$ latency speedup and $3.68\times$ energy savings compared to SOTA accelerators, without accuracy degradation.

\end{abstract}



\maketitle

\section{Introduction}
\label{sec:intro}


The rapid advancement of 
artificial general intelligence (AGI) 
empowers AI systems to generate diverse and 
realistic content.
Recently, 
Diffusion Transformers (DiTs) have achieved state-of-the-art performance across various tasks, including multimedia synthesis \cite{SDXL}, \cite{PIXART}, \cite{SVD}, \cite{OPENSORAPLAN}, content modification \cite{IMAGEEDIT}, and super-resolution \cite{SUPERRES}.
However, DiTs face significant inefficiency
due to the multi-iteration inference process required by large-scale models. 
For example, Flux-1 \cite{FLUX}, a cutting-edge image generation model with 12 billion parameters, requires 924 TOPs to generate a 1K-resolution image in a 28-step inference process, consuming 25 seconds and 7475 J on an Nvidia A100 GPU. 

To improve efficiency, 
various algorithmic optimizations have been explored.
Some approaches \cite{DDIM,DPMSOLVER,DITDISTILLATION} aim to reduce the number of timesteps in the diffusion process through stochastic methods and distillation techniques. 
Methods like \cite{DEEPCACHE, TOKENCACHE, CACHESERVER, MODM, PATCHSERVER, FLEXCACHE} cache intermediate results of previous steps for reuse and reduce computation across scenarios ranging from edge to server-side deployment.
While these approaches accelerate inference, they often lead to great degradation in the quality of generated content when applied to DiTs. For example, TokenCache~\cite{TOKENCACHE} achieves a 2× speedup on A100 GPU but results in over a 5\% drop in FID.

Recent studies have explored software-hardware co-design strategies to accelerate the diffusion process with minimal degradation in model performance. In particular, activations generated at adjacent timesteps often exhibit high similarity, revealing substantial \textit{temporal redundancy}. For linear layers, since the differences between adjacent timesteps, referred to as \textit{differential activations}, are typically small, they can be quantized to lower precision (e.g., 4 bits) \cite{CIMDIFFACC,CAMBRICOND,DITTO,EDGEDIFF}, or even set to zero \cite{EXION}, thereby reducing computational cost. 
Meanwhile, exploiting \textit{spatial redundancy}, i.e., sparsity in attention layers, has received significant attention. 
Prior accelerators for transformers like BERT \cite{BERT} leverage this sparsity by predicting important attention scores and skipping redundant computations \cite{SOFA,FACT,SANGER,ELSA}, thereby reducing the computational burden.




\begin{figure}[]
    \centering
    \includegraphics[width=\columnwidth]{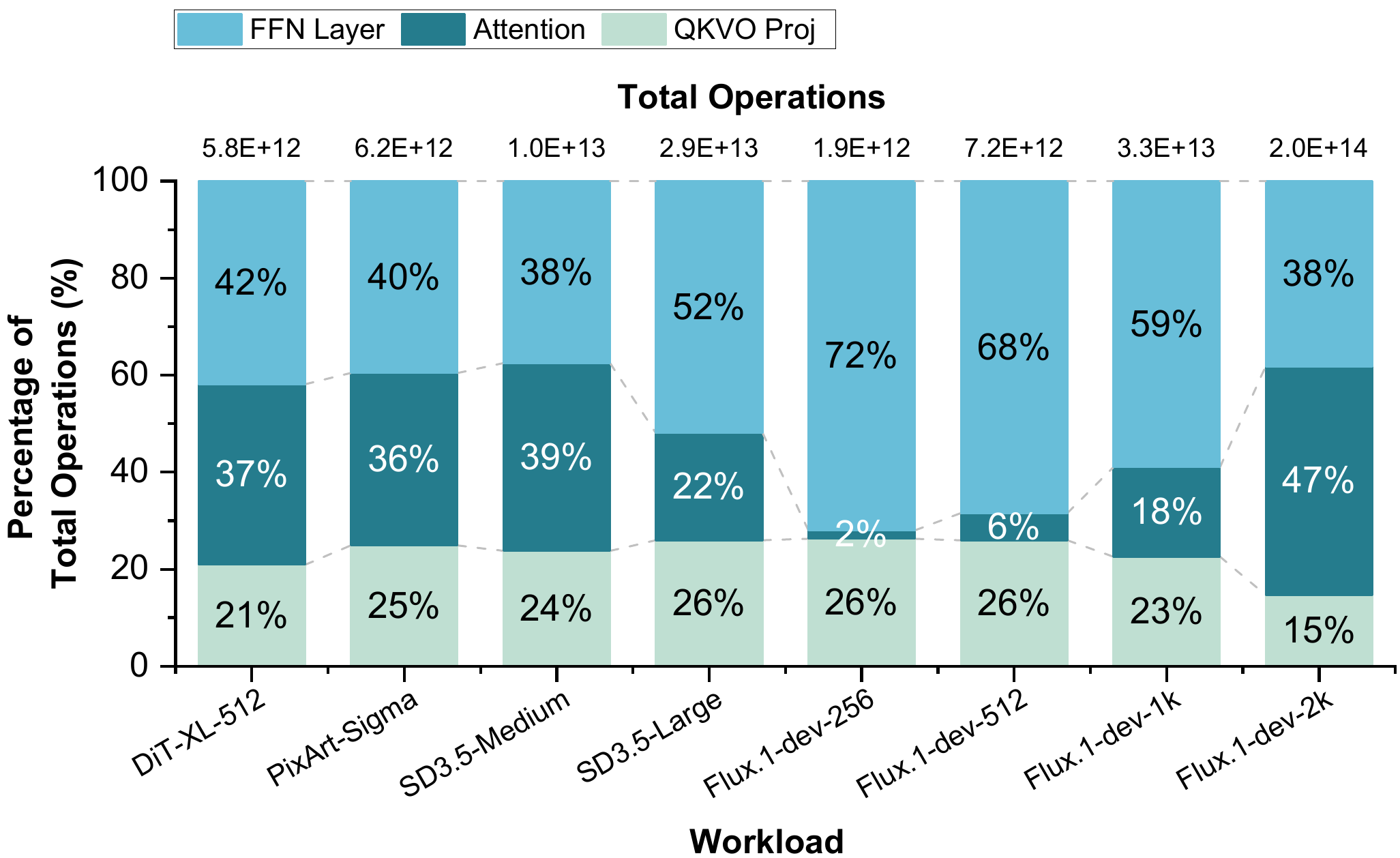} 
    \caption{Performance breakdown for different DiT models.}
    \label{fig:workload_breakdown}
\end{figure}

\textbf{Challenges.} DiTs inherently possess characteristics of both the diffusion process and transformer models, creating substantial opportunities to exploit temporal and spatial redundancies. However, fully leveraging these redundancies in DiTs is non-trivial due to several key challenges. 
For linear operations such as GEMM in FFNs, existing differential-based methods \cite{CAMBRICOND, DITTO} employ coarse-grained outlier-aware quantization to quantize differential activations. However, such coarse-grained schemes introduce high quantization errors when applied to DiTs, limiting the ability to compress large tensors into low-bit representations. Effectively quantizing as many differential activations as possible to low precision without compromising generation quality remains a significant challenge.

For attention operations, although sparse attention techniques have been explored in both traditional Transformers \cite{SOFA, FACT, SANGER, ELSA} and DiT models \cite{EXION, SPARGEATTN}, their prediction-based approaches introduce considerable overhead and fail to exploit the distinctive attention patterns of DiT.
Moreover, temporal redundancy also exists within attention layers of DiTs. Existing methods either suffer from accuracy degradation \cite{TOKENCACHE} or are limited to specific attention heads \cite{DITFASTATTN}. Thus, effectively leveraging this redundancy presents a valuable yet underexplored opportunity.


\textbf{Our Solution.} The performance bottlenecks in DiTs vary significantly depending on the generation task, model scale, and input resolution, as shown in Fig. \ref{fig:workload_breakdown}. In different DiTs, either the attention or the FFN may dominate inference latency. This variation highlights the need for a comprehensive acceleration strategy that targets all major components of DiTs. We address it by reducing both spatial and temporal redundancy, and propose a co-design approach that integrates algorithmic and hardware optimizations.

At the algorithm level, we introduce a fine-grained mixed-precision quantization method tailored for differential activations of linear operations like FFN, significantly increasing the proportion of values that can be quantized to lower bits. Moreover, we propose a sparse attention reuse mechanism for attention operations, performing block-wise sparse computation at each timestep and reusing attention scores from previous timesteps. Our algorithms reduce redundancy of DiTs without degrading generation quality.


For architectural support, we incorporate a specialized hardware architecture which extends the TPU design by redesigning the functionality and arrangement of PEs and adding additional functional units. This architecture is co-designed with our algorithmic optimizations, enabling high execution efficiency while preserving generation quality.
Our contributions are summarized as follows:
\begin{itemize}
    \item We propose DSTAR, a software-hardware co-design framework that accelerates DiT inference by reducing both spatial and temporal redundancy.
    \item We introduce a fine-grained differential quantization algorithm for linear operations and a sparse attention reuse mechanism for efficient attention computation.
    \item We provide a high-performance hardware accelerator to support our algorithmic optimizations.
\end{itemize}

We evaluate DSTAR on seven typical DiT models spanning various generative tasks. Experiments show that 
DSTAR achieves up to $7.33\times$ latency speedup and $41.89\times$ energy savings compared to an NVIDIA A100 GPU, and achieves up to $2.54\times$ latency speedup and $3.68\times$ energy savings compared to state-of-the-art accelerators, without accuracy degradation.



\section{Background and Motivation}
\label{sec:background}

\begin{table*}[]
  \centering
  \caption{Comparison of representative accelerators.}
  \resizebox{\textwidth}{!}{%
    \begin{tabular}{c|cccc|cccc}
    \toprule
    \multirow{2}[4]{*}{\textbf{Accelerators}} & \multicolumn{4}{c|}{\textbf{Differential Quantization for Linear Layer}} & \multicolumn{4}{c}{\textbf{Pruning for Attention Layer}} \\
\cmidrule{2-9}          & \textbf{Granularity} & \multicolumn{1}{c}{\textbf{Precision Support}} & \makecell{\textbf{Avg.} \\ \textbf{Bit-width}} &\makecell{\textbf{HW} \\ \textbf{Friendly}} & \textbf{Granularity} & \textbf{Overhead} & \textbf{Reuse} & \makecell{\textbf{Pruned} \\  \textbf{Ratio} } \\
    \midrule
    Cambricon-D \cite{CAMBRICOND} & Channel-wise & \makecell{Low Magnitude : INT4\\High Magnitude : FP16}  & 7.2   & \ding{51}  & $-$   & $-$ & \ding{55} & 0\% \\
    \hline
    EXION \cite{EXION} & Element-wise & \makecell{Low Magnitude : Set to 0\\High Magnitude : FP16} & 11.2  & \ding{55} & Top-K Per-Q  & High & \ding{55} & 17\% \\
    \hline
    DITTO \cite{DITTO} & Channel-wise & \makecell{Low Magnitude : INT4\\High Magnitude : INT8} & 5.4   & \ding{51}  & $-$    & $-$ & \ding{55} & 0\% \\
    \hline
    Ours  & Sub-channel-wise & \makecell{Adaptive : INT2 / INT4 \\ / INT6 / INT8}  & \textbf{4.3}   & \ding{51}  & Block-wise    & Low & \ding{51}  & \textbf{37\%} \\
    \bottomrule
    \end{tabular}%
    }
  \label{tab:comparison}%
\end{table*}%

\subsection{Diffusion Transformer}

\begin{figure}[]
    \centering
    \includegraphics[width=0.98\columnwidth]{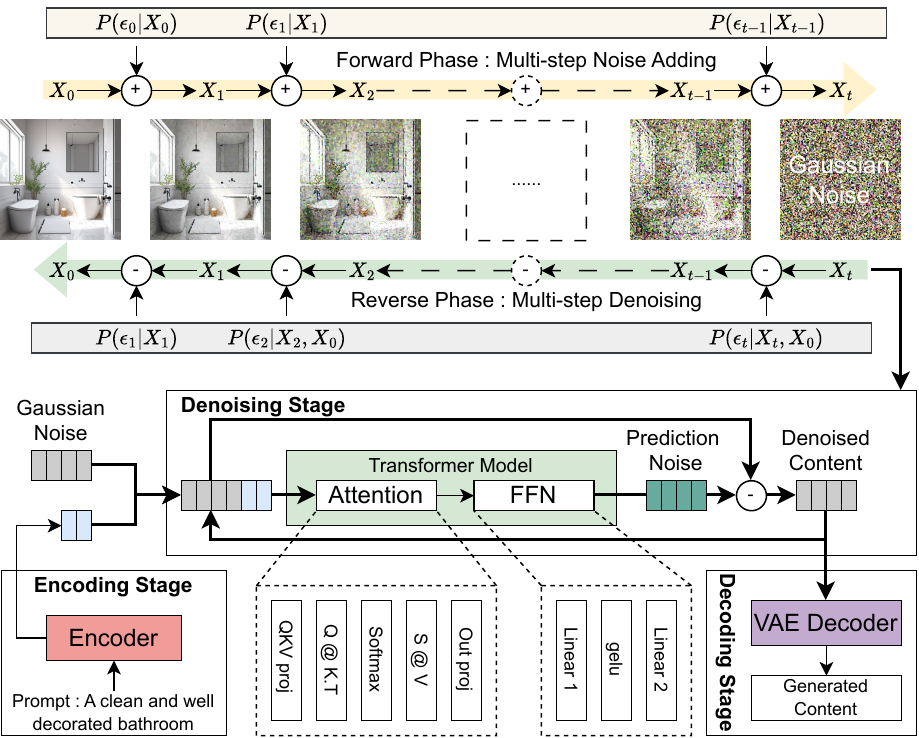} 
    \caption{The basic process of diffusion transformer.}
    \label{fig:dit_background}
\end{figure}

Diffusion is a generative process designed to sample from a target data distribution, such as images or videos, enabling the generation of corresponding content. As shown in Fig. \ref{fig:dit_background}, taking the example of image generation, the diffusion process is formulated as a Markov process with a forward phase and a reverse phase. 



In the forward phase, a sequence of Gaussian noise perturbations \(\epsilon_t\) is progressively added to the original image \(X_0\), generating a sequence of images, \(X_0, X_1, \dots, X_n\). When \(n\) is sufficiently large, \(X_0\) is transformed into pure Gaussian noise, i.e., \(X_n \sim \mathcal{N}(0, I)\).
Mathematically, when the noise added in the forward phase is small enough, the conditional probability of the inverse noise \(P(\epsilon_{t} | X_t, X_0)\) can be well approximated by the forward noise distribution \(P(\epsilon_{t-1} | X_{t-1})\). This property enables the restoration of the original data distribution by iteratively subtracting noise from the Gaussian distribution, as shown in the reverse denoising phase.

The generation pipeline is depicted in
Fig. \ref{fig:dit_background}. Conditioning inputs, such as class labels, text prompts, or images, are first encoded using 
specialized encoders, such as CLIP \cite{CLIP} for vision-language tasks or T5 \cite{T5} for text processing. These encoded conditions, along with an initial Gaussian noise sample, serve as inputs to the noise prediction model. The denoising phase is then executed, iteratively refining the noisy input through multiple denoising steps based on the model's prediction. For models operating in a latent space, a variational autoencoder (VAE) is employed at the final stage to decode the latent representation into generated content.

The denoising model is the core of the diffusion process and dominates inference latency due to the need for multiple iterations. Specifically, DiTs have become the de facto choice for recent diffusion-based applications. A DiT typically consists of several Transformer blocks, each comprising an attention layer and a FFN layer. Both layers involve several matrix multiplications and nonlinear arithmetic operations.

\textbf{Performance Breakdown.} Fig. \ref{fig:workload_breakdown} presents a performance breakdown of main computational layers across different workloads and different DiT models. It indicates that \textit{the bottlenecks in DiTs vary depending on the task (image vs. video generation), model scaling (small vs. large models), and input size (image and video resolution)}. Both attention and FFN layers could dominate the inference latency, highlighting the necessity of accelerating all major components. 

Existing accelerators are primarily designed for UNet \cite{CAMBRICOND, DITTO, SDMA} or small-scale DiTs with fewer layers and lower generation resolutions such as DiT-XL \cite{DITTO, EXION}, 
and lack the flexibility to efficiently support a wide range of DiTs (Sec. \ref{sec:temporal-linear} and Sec. \ref{sec:redund-attention}). To address this issue, we accelerate both linear and non-linear operations in DiT by eliminating spatial and temporal redundancy, and introduce an efficient software-hardware co-design framework to enhance performance of diverse DiT models.  
\subsection{Temporal Redundancy for Linear Layers}
\label{sec:temporal-linear}

\textbf{Temporal Redundancy.} 
Activations generated in adjacent steps often exhibit high similarity due to minimal noise addition, 
indicating high \textit{temporal redundancy} \cite{CAMBRICOND, DITTO}.


To reduce temporal redundancy, \textit{differential-based inference} 
is applied to linear layers
such as FFN.
Consider a linear transformation such as \( y = kx + b \), where \(k\) is constant. When an increment \( \Delta x \) is applied to \( x \), the updated value \( y' \) can be computed using the linearity property: 
\begin{equation}
\label{equal-diff}
    y' = y + \Delta y = y + k \Delta x.
\end{equation}
Since \( \Delta x \) is typically small, it can be either quantized to fewer bits, or directly set to zero, 
thus simplifying computation.

Previous research have leveraged this idea to accelerate diffusion models. For example, EXION \cite{EXION} directly sets differential activations of FNN layers to zeros,
achieving good performance on UNet. 
However, as shown in Table~\ref{tab:comparison}, it achieves an average bit-width of 11.2 when applied to DiT models. Moreover, it introduces additional hardware overhead due to the need of processing tensors that contain unstructured zeros produced by their algorithm.
\textit{Differential-based quantization} is a superior alternative to reduce temporal redundancy for UNet \cite{CAMBRICOND, CIMDIFFACC} and DiT \cite{DITTO}, 
and selecting an appropriate quantization granularity is crucial to balance accuracy and efficiency.
Similar to OliVe \cite{Olive} and MSQ \cite{MSQ} for LLM, Cambricon-D~\cite{CAMBRICOND} adopts a coarse-grained outlier-aware quantization strategy for UNet that partitions activations into two groups---outliers and normal values---and quantizes them separately. 
However, it introduces high quantization error when applied to DiTs.
DITTO~\cite{DITTO} supports Per-Channel Quantization (PCQ)~\cite{QDIFF, PCQ, GPT3INT8, TENDER, OLTRON} for DiT, which decomposes activations along the channel dimension into multiple sub-tensors (each contain a 1-D channel), each quantized with its own precision and scale.
However, our experiments show that PCQ still introduces substantial quantization error when applied to the differential activations of large-scale DiT models such as Stable Diffusion 3.5 \cite{SD3} and Flux.1-Dev \cite{FLUX}, resulting in a much higher average bit-width requirement, as reflected in the average bit-widths of DITTO in Table~\ref{tab:comparison}. 
\textbf{Observation and Motivation.} 
To quantize differential activations to low bit-widths without incurring significant accuracy degradation, we first analyze their outlier distribution patterns, as shown in Fig. \ref{fig:outlier_distribute}. 
\textit{In general, outliers are dispersed across channels, but within each channel they exhibit strong clustering. This motivates the use of finer-grained quantization that exploits the intra-channel locality of outliers, enabling both higher efficiency and better generation quality. }
However, such fine granularity poses additional challenges for hardware efficiency. To reconcile this trade-off, we propose a fine-grained mixed-precision differential quantization algorithm together with a hardware architecture tailored to support it. Details are provided in Sections~\ref{sec:alg-quant} and~\ref{sec:hardware}.

\subsection{Spatial-Temporal Redundancy for Attention}\label{sec:redund-attention}

\textbf{Spatial Redundancy.}
Sparse attention techniques accelerate attention computation by reducing \textit{spatial redundancy}. 
Due to the nature of the softmax, which pushes small values toward zero, the attention score matrix naturally becomes sparse. 
Some methods employ dynamic prediction to estimate the sparsity pattern of the score matrix at runtime for both traditional Transformers \cite{SOFA, FACT} and DiTs \cite{SPARGEATTN, EXION}. 
For example, as shown in Table~\ref{tab:comparison}, EXION adopts an approximate top-\(k\) strategy that predict the largest \(k\) values for each query and sets the rest to zero.
However, such prediction introduces additional latency, area, and power overhead. 
Alternatively, static sparsity patterns---such as attention sinks, window attention, and global attention \cite{DIGPATTERN, STREAMLLM, WINDOWATTN}---are widely used in traditional Transformers to reduce computation. 
Unfortunately, they are not directly applicable to DiTs, which exhibit distinct sparsity characteristics.
In addition, some works \cite{SDMA} explore token-pruning methods 
for specific models and tasks, such as image editing; however, they 
do not generalize well to generative workloads in DiTs.

\textbf{Temporal Redundancy.} \textit{Temporal redundancy} also exists in attention operations within DiTs, stemming from the temporal similarity between adjacent steps. As with linear layers, differential-based acceleration can in principle be applied to attention. However, its effectiveness is limited because differential activations in attention layers are more sensitive to quantization errors, degrading model performance greatly. Consequently, direct activation reuse is more suitable for attention. The key challenge lies in determining which values can be safely reused across steps without compromising generation quality.
For instance, some approaches reuse attention output tokens but incurs substantial accuracy degradation \cite{TOKENCACHE}.
DiTFastAttn \cite{DITFASTATTN} observes that reusing attention outputs for heads exhibiting a diagonal pattern introduces minimal error. Nonetheless, it is restricted to heads with such patterns, which are relatively scarce in DiTs. 



\textbf{Observation and Motivation.} 
To reduce spatial and temporal redundancies for attention, we identify a simple yet effective sparsity pattern in DiTs, as shown in Fig. \ref{fig:sparsity-partten}. 
In particular, \textit{a block-wise sparsity pattern emerges within each score matrix, reflecting spatial structure, while this pattern exhibit similarity across steps,
capturing temporal correlation.} 
Therefore, we design a pattern-aware strategy that performs block-wise sparse computation at each step and reuses \textit{attention scores} from the previous step, as detailed in Sections \ref{sec:alg-sparse} and \ref{sec:hardware}.
Notably, 
we observe that the L1 norm of the differences in attention components between adjacent steps exhibits vastly different value ranges, indicating that they have different tolerances to reusing error.
For example, the value matrix
changes significantly between steps. In contrast, due to the normalization effect of softmax, attention scores are compressed into a narrow range \([0, 1]\), making them less sensitive to final result and hence
a suitable target for exploiting temporal redundancy in attention.







\begin{figure}[]
    \centering
    \includegraphics[width=\columnwidth]{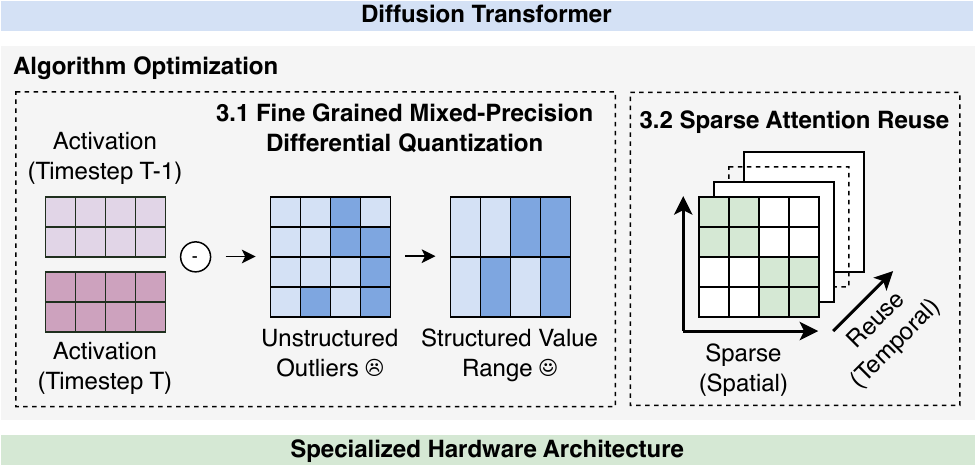} 
    \caption{Overview of the DSTAR algorithmic optimization.}
    \label{fig:alg_overview}
\end{figure}

\section{Algorithm Optimization}
\label{sec:alg}

To fully exploit spatial and temporal redundancies in DiTs, we propose two algorithmic optimizations that target linear layers like FFN and attention layers, respectively, as shown in Fig. \ref{fig:alg_overview}. We first introduce a fine-grained mixed-precision differential quantization (FMDQ) scheme to improve hardware efficiency while maintaining accuracy.
Next, we propose a sparse attention reuse (SAR) scheme that exploits spatial sparsity and temporal similarity in attention.

\subsection{Fine-Grained Mixed-Precision Quantization}
\label{sec:alg-quant} 
\begin{figure}[]
    \centering
    \includegraphics[width=\columnwidth]{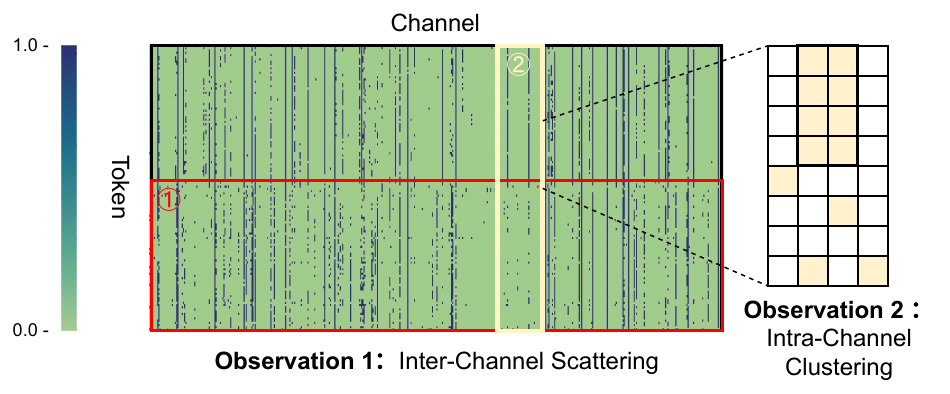} 
    \caption{Distribution of outliers in differential activations.}
    \label{fig:outlier_distribute}
\end{figure}

\textbf{Key Idea.}  Take the Stable Diffusion 3.5 Medium model \cite{SD3} as an example. Fig. \ref{fig:outlier_distribute} shows the binary heatmap of differential activation values of layer 23 between step 0 and step 1, indicating the position of outliers in the tensor. We can obtain two key observations. 
First, \textit{outliers are unevenly distributed across most channels}, posing a significant challenge for coarse-grained quantization scheme like PCQ  \cite{QDIFF, PCQ,GPT3INT8, TENDER, OLTRON}, which is originally designed for cases where outliers are concentrated within specific channels. 
For the model in Fig. \ref{fig:outlier_distribute}, applying PCQ means that even if a channel contains only a few outliers, the entire channel must still be processed at high precision, substantially reducing quantization efficiency. 
Second, \textit{outliers within each channel often cluster among adjacent tokens}, presenting an opportunity for a finer-grained quantization strategy, namely, separating outliers from normal values \textit{within each channel} to enable more efficient computation.

\begin{figure*}[]
\centering
\includegraphics[width=\textwidth]{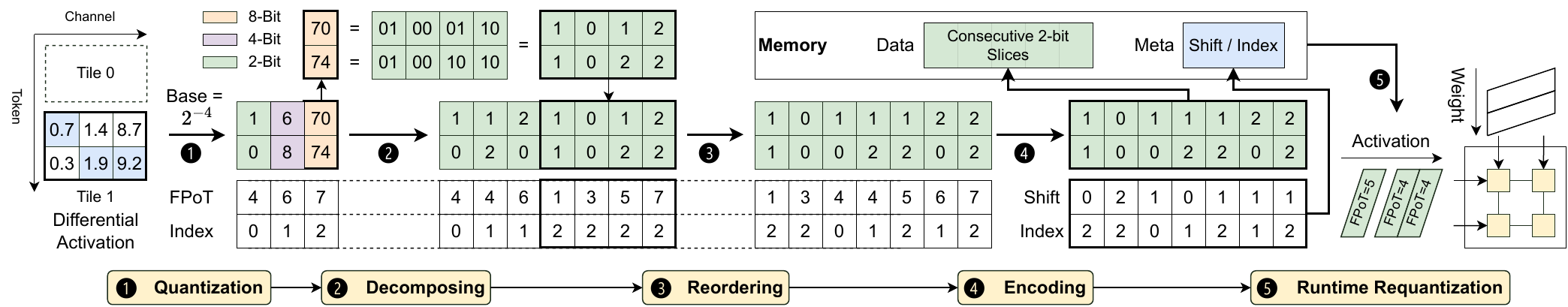}
\caption{Computation flow of fine-grained mixed-precision differential quantization.}
\label{fig:computation_flow}
\end{figure*}

\textbf{Computation Flow.} Fig. \ref{fig:computation_flow} illustrates the computation flow of our FMDQ method. 
First, we divide the differential activation matrix into fine-grained tiles along the token dimension, 
and then apply an on-the-fly power-of-two (PoT)-based quantization scheme for each channel of each tile, i.e., \textit{\textbf{sub-channels}} (\ding{182}).
Next, we decompose mixed-precision sub-channels into several slices represented with a unified precision (\ding{183}). 
Subsequently, we reorder sub-channels based on their PoT so that those sharing the same PoT are placed adjacently, thereby reducing stalls and improving hardware efficiency (\ding{184}). 
After that, the number of required shifting operations are calculated based on PoTs, and we encode the data along with meta-information into a compact format for memory efficiency (\ding{185}). Finally, the quantized data is streamed to the computation module of our specialized accelerator, where runtime requantization is conducted to enable efficient partial sum accumulation (\ding{186}).


\textbf{\ding{182} On-the-Fly PoT-Based Quantization.} 
Based on our observation from Fig. \ref{fig:outlier_distribute}, we divide the tensor along the token dimension into several tiles, since outliers usually cluster to adjacent tokens. 
As shown in Fig. \ref{fig:computation_flow}, the colored and blank tensors represent two tiles. 
Within each tile, we apply mixed-precision quantization at the per-channel granularity, assigning INT2, INT4, INT6, or INT8 precision depending on the value range of each channel. 
The reason for supporting 2-bit incremental precision is that the differential activations in DiT exhibit significant variation. Across different steps, the first and last steps tend to change more sharply, whereas intermediate steps show relatively small differences. Moreover, the number of inference steps varies substantially across DiT models: some models require more than 50 steps, introducing considerable redundancy, while others need only around 4 steps, where activation values change much more rapidly. 
With fine-grained mixed-precision support, FMDQ can effectively adapt to a wide range of DiT models and inference configurations.




To efficiently determine the precision and scale for each sub-channel and perform quantization, we propose an \textit{on-the-fly} PoT-based linear symmetric quantization scheme.
PoT quantization \cite{POT} is an efficient method where quantization scales are restricted to powers of two relative to a base scale, allowing requantization to be efficiently implemented using bit-shift operations on hardware.

Let the maximum absolute value of each sub-channel be denoted as \(M_i\). We first determine the quantization precision by binning \(M_i\) (i.e., assigning a value to a range) using predefined thresholds \(Thr_0, Thr_1, \ldots\). These thresholds are defined as a series \(Thr_0 \times 2^{2N}\), where each threshold is \(2^2\) times larger than the previous one, corresponding to the value ranges representable by different precisions (e.g., 2-bit, 4-bit, etc.). By tuning \(Thr_0\), we can encourage the selection of lower precision while maintaining negligible accuracy loss, as demonstrated in Sec.~\ref{sec:eval-alg}.

Let the maximum representable value for the selected precision be denoted as \(PCeil\).
Our PoT-based quantization selects the minimum of \(M_i\) across all sub-channels as \(base\), and determines quantization scales $s_i$ of each sub-channel with:
\begin{equation}
    s_i = \frac{PCeil}{base \times 2^{PoT_i}}
    \label{eq:scale}
\end{equation}

where
\begin{equation}
    PoT_i = \left\lceil \log_2\!\left(\frac{M_i}{base}\right) \right\rceil.
    \label{eq:pot}
\end{equation}

For convenience, we further define
\begin{equation}
    FPoT_i = \log_2\!\left(\frac{PCeil}{base^2 \times 2^{PoT_i}}\right),
    \label{eq:fpot}
\end{equation}
which fuses the information of precision ($PCeil$) and PoT value ($PoT_i$), serving as a unified indicator 
for requantization across quantized values with different precisions and scales.  
Accordingly, the scale can then be simplified as
\begin{equation}
    s_i = base \times 2^{FPoT_i}.
    \label{eq:si_fpot}
\end{equation}

The PoT-based quantization 
relies on finding \( M_i \) and \( base \), where \( M_i \) is the maximum value in a certain sub-channel (usually 64 to 256 elements). This is easy to store in on-chip memory. However, determining \( base \) requires finding the global minimum of \( M_i \) across all sub-channels.
This process doubles memory accesses and latency, compared to non-PoT quantization which depends only on local value ranges, as it requires at least two passes over the data.

Our key insight is that using a predefined static \(base\), selected through dataset profiling, eliminates the need to compute the global minimum at runtime.  Once \(M_i\) of a sub-channel is obtained, a threshold-based method can be applied 
to determine the quantization precision (2-bit, 4-bit, 6-bit, or 8-bit) for that sub-channel.  
Subsequently, \(PoT\), \(FPoT\) and the quantization scale can be directly derived from 
Eq.~\ref{eq:pot}, 
Eq.~\ref{eq:fpot} and Eq.~\ref{eq:si_fpot}, thereby enabling efficient 
on-the-fly quantization and meta-information generation.
Experiments demonstrate that this approach achieves accuracy
comparable to that of using the dynamic global minimum.
Moreover, since both the \(base\) and thresholds for different precisions are fixed, 
the PoT and precision can be efficiently determined by classifying values into predefined ranges, i.e., through a bin operation.  
For example, consider the first sub-channel of the tensor in Fig.~\ref{fig:computation_flow}\ding{182}, where the maximum absolute value is \(M_1 = 0.7\). 
Suppose the threshold for 2-bit quantization is set to \(2^{0}\).  
Since \(M_1\) lies within the range \([0,\,2^{0}]\), it is assigned to 2-bit quantization. 
In this case, $PCeil=1$. 
Given that $base$ is set to $2^{-4}$ in this example, we can obtain \(PoT = 4\), \(FPoT = 4\), and \(s_i = 1\) with Eq.~\ref{eq:pot} to Eq.~\ref{eq:si_fpot}.
Note that all values in the interval \([0,\,2^{0}]\) share the same FPoT and scale. 
These parameters can be directly retrieved from a precomputed lookup table.

Once the quantization scale obtained, all the values in the sub-channel can be quantized to the corresponding precision. 

\textbf{\ding{183} Sub-Channel Decomposition.} 
Handling mixed-precision values introduces hardware inefficiency due to the imbalanced utilization of PEs
~\cite{OLTRON}. This problem is further amplified in the diffusion process, where activation value ranges vary significantly across steps and layers.

To simplify the hardware architecture and improve efficiency, we decompose each sub-channel into several unified 2-bit slices. Consider a sub-channel \(A\) consisting of 64 elements, each represented in 8 bits. This sub-channel can be expressed as  
\[
A = (\text{Slice[3]} \ll 6) + (\text{Slice[2]} \ll 4) + (\text{Slice[1]} \ll 2) + \text{Slice[0]},
\]  
where \(\text{Slice}[n]\) denotes a 64-element vector containing the \(n\)-th 2-bit component of A. We then replace \(A\) with multiple slices.  
However, a direct replacement leads to incorrect results for matrix multiplication. For example,  
\[
A \cdot B = \big((\text{Slice[1]} \ll 2) + \text{Slice[0]}\big) \cdot B \neq \text{Slice[1]} \cdot B + \text{Slice[0]} \cdot B.
\]  
To address this, we adjust the \(FPoT\) as follows: no adjustment is required for \(\text{Slice[0]}\), while for \(\text{Slice[1]}\), \(\text{Slice[2]}\) and \(\text{Slice[3]}\), \(FPoT\) values must be decreased by 2, 4, and 6, respectively. This ensures that during the following partial-sum accumulation, each slice is effectively shifted to its correct position. An example of this decomposition is illustrated in Fig.~\ref{fig:computation_flow}\ding{183}, where the third sub-channel with values 70 and 74 is decomposed into four separate slices following the described method.

\begin{figure}[]
    \centering
    \includegraphics[width=\columnwidth]{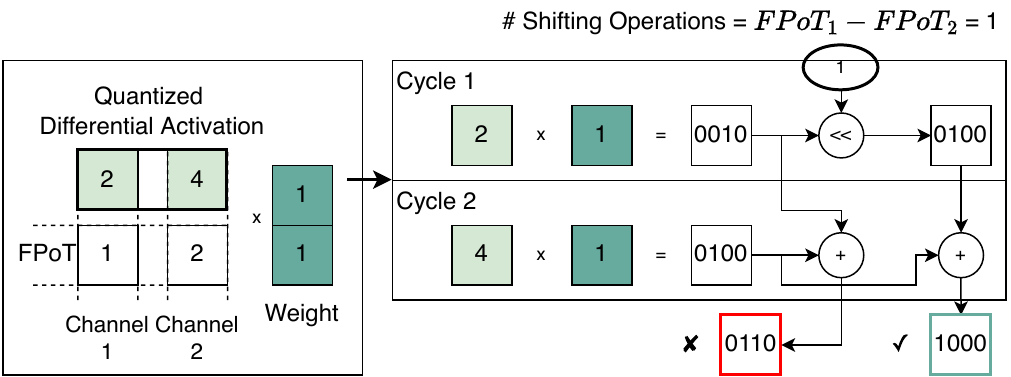} 
    \caption{Illustration of requantization: shifting operations are required when accumulating partial sums.}
    \label{fig:requant}
\end{figure}

To validate correctness, we take the value 70 in Fig.~\ref{fig:computation_flow}\ding{183} as an example, 
which is decomposed into 1 (Slice[3]), 0 (Slice[2]), 1 (Slice[1]), and 2 (Slice[0]), 
with corresponding \(FPoT\) values of 1, 3, 5, and 7, respectively.  
To compute \(70 \times w\), we perform four iterations: 
first, compute \(1 \times w = w\) (i.e., $Slice[3]\times w$); 
second, compute \((w \ll 2) + 0 \times w = 4w\) 
(i.e., $+=Slice[2]\times w$).  
The left shift by 2 corresponds to a requantization step (Fig. \ref{fig:requant}), derived from the FPoT difference 
\(\Delta \text{FPoT} = 3 - 1 = 2\).  
This process continues for the remaining slices, and we can ultimately verify that 
the correct result \(70w\) is obtained.


\textbf{\ding{184} Sub-Channel Reordering, \ding{185} Encoding and \ding{186} Requantization.}
After unifying precision, it is necessary to cluster sub-channels with identical 
FPoTs to enable efficient runtime requantization \cite{TENDER}. As shown in Fig.~\ref{fig:requant}, consider two quantized differential activations that are multiplied by weights to produce partial sums ($0010$ and $0100$). If their FPoTs differ, indicating different quantization scales, shifting operations are required to align them before accumulation. This process, known as \textit{requantization}, involves adjusting partial sums by the difference in FPoT to ensure correctness.

Our key idea is to group sub-channels with identical FPoTs and reorder them based on FPoT magnitudes, thereby minimizing the number of shifting operations during partial-sum accumulation.
We propose a reordering scheme (\ding{184}) in which all 2-bit slices, together with their FPoT and indices, are sorted using a bitonic sorting network. Thanks to the high parallelism of the on-chip sorter in our architecture, the entire sorting process can be performed on-the-fly, making it highly efficient.  
After reordering, multiplications and partial-sum accumulations are executed in ascending order of FPoT, which effectively reduces the requantization cost.



During encoding (\ding{185}), FPoT values are converted into the number of shifting operations, calculated as the difference between adjacent FPoTs. The quantized data, along with its meta-information, including channel indices and the number of shifts, is encoded and stored in consecutive memory locations.   
Channel indices are used to retrieve the correct weights 
during computation. All the data and its meta-information are streamed into the computation core of DSTAR, enabling efficient runtime requantization (\ding{186}).

\subsection{Sparse Attention Reuse}
\label{sec:alg-sparse}

\begin{figure}[]
    \centering
    \includegraphics[width=\columnwidth]{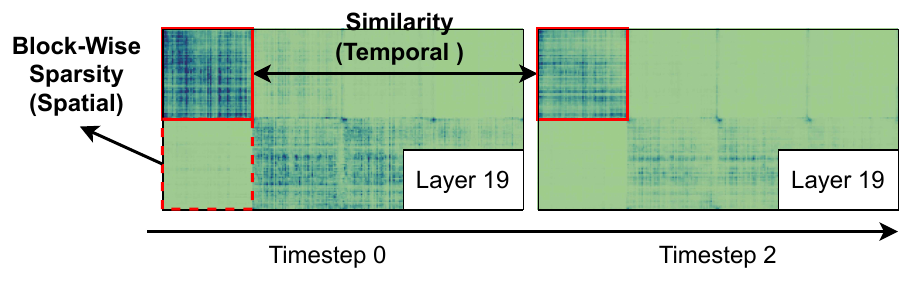} 
    \caption{The sparsity pattern of the attention score matrices at different steps for the SD3.5-Medium model \cite{SD3}.}
    \label{fig:sparsity-partten}
\end{figure}

We identify a simple yet effective sparsity pattern, as shown in Fig. \ref{fig:sparsity-partten}. A \textit{block-wise} sparsity pattern appears in each score matrix, demonstrating \textit{spatial redundancy}, and these sparse blocks consistently appear in the same positions of the score matrix, regardless of input variations. 
We attribute this to DiTs learning inductive biases aligned with the structure of image or video data. 
Across different timesteps, the sparsity patterns exhibit notable \textit{temporal similarity} at the same layer between adjacent steps.
Therefore, the block-wise sparse attention scores can be reused, 
reducing computational overhead. However, this approximation introduces accumulated errors across timesteps. To mitigate this, we periodically recompute attention scores every $n$ steps to maintain model accuracy.

\begin{figure}[]
\centering
\includegraphics[width=\columnwidth]{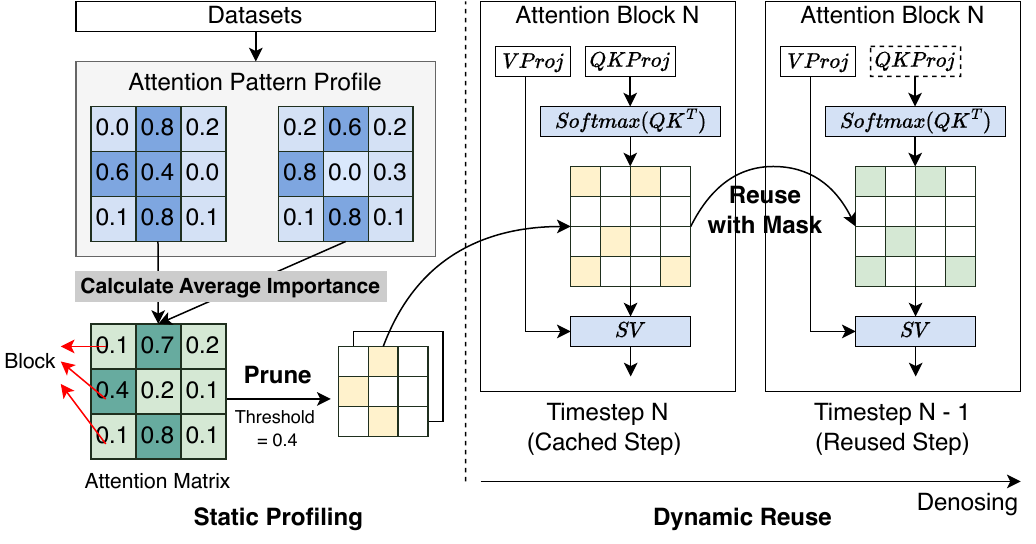}
\caption{Illustration of sparse attention reuse. 
}
\label{fig:workflow_sar}
\end{figure}


\textbf{Approach}. Based on our observations, we propose Sparse Attention Reuse (SAR), a pattern-aware strategy designed to accelerate attention in DiTs. SAR performs block-wise sparse computation at each step (spatial) and reuses attention scores from the previous step (temporal). 
The workflow, shown in Fig. \ref{fig:workflow_sar}, consists of two stages: static profiling and dynamic reuse. During profiling, the model is executed on various inputs from the dataset, using different random seeds, and attention scores are collected. Based on observed patterns, the score matrix is divided into blocks 
and the statistical importance of each block is computed by summing all attention scores within it. As illustrated in Fig.~\ref{fig:workflow_sar}, these block-wise importance values are then averaged to obtain the final importance score. Then blocks with importance score that fall below a predefined threshold are pruned, generating a block-wise sparse attention mask which will guide computation during inference. 
Experiments show that the profiling process is highly dependent on model weight and the number of steps. Each time one of these factors changes, re-profiling is required. However, in practical scenarios, they are typically fixed, and the profiling incurs only minimal overhead, making it affordable and practical.

During inference, we employ a cache-and-reuse mechanism. At the initial step (timestep \(N\) in Fig.~\ref{fig:workflow_sar}), referred to as the cached step, we compute the sparse attention matrix and store it in the cache. In the subsequent \(n\) steps, termed reused steps, we leverage the cached score matrix to bypass the computation of the \(Q\) and \(K\) projections, as well as \( \mathrm{softmax}(QK^T) \). This caching and reuse process continues throughout the remainder of the inference.




\section{Hardware Architecture}
\label{sec:hardware}
\subsection{Overview}

\begin{figure}[]
    \centering
    \includegraphics[width=0.95\columnwidth]{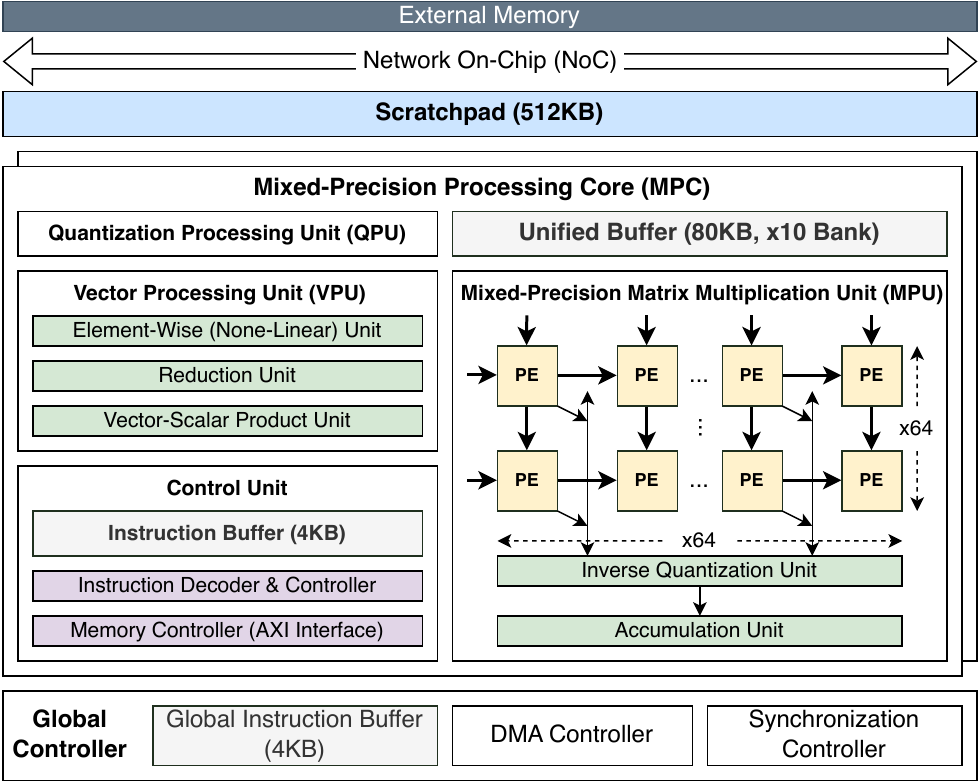} 
    \caption{Hardware architecture of \ACCELERATOR. 
    }
    \label{fig:arch_overview}
\end{figure}

Fig. \ref{fig:arch_overview} illustrates the overall hardware architecture. DSTAR adopts a multi-core design composed of multiple mixed-precision processing cores (MPCs), a global controller, and a scratchpad memory. All components are interconnected through the network on-chip (NoC), which interfaces with external memory.


\subsection{Mixed-Precision Processing Core (MPC)}

Each MPC consists of several specialized processing units,
including a mixed-precision matrix multiplication unit (MPU) for efficient tensor computation, a quantization processing unit (QPU) for our fine-grained quantization, and a vector processing unit (VPU) for other operations such as GELU and Softmax. 
The control unit orchestrates the dataflow based on instructions and manages memory accesses, interfacing with scratchpad and external memory.

\begin{figure}[]
    \centering
    \includegraphics[width=0.97\columnwidth]{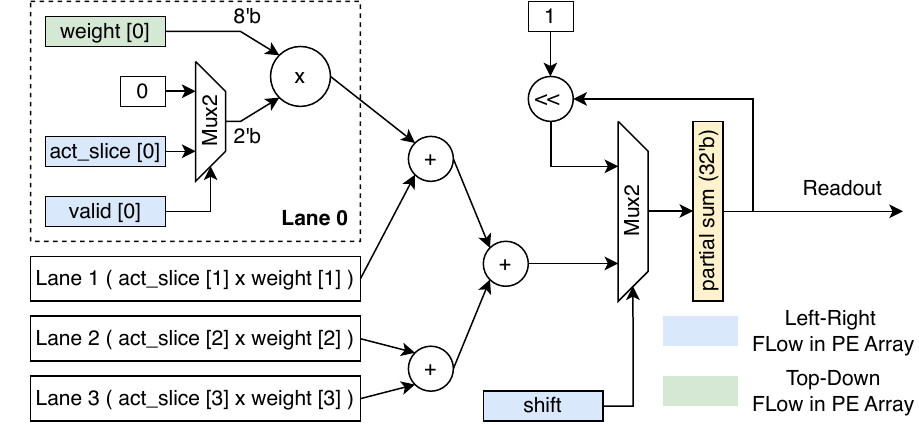} 
    \caption{Design of PE in MPU.}
    \label{fig:pe}
\end{figure}

\textbf{Design of MPU}. As shown in Fig. \ref{fig:arch_overview}, MPU consists of a $64 \times 64$ PE array, an inverse quantization unit, and an accumulation unit. Fig. \ref{fig:pe} illustrates the microarchitecture of a PE in MPU. Each PE performs a SIMD-4 inner product between a 2-bit activation slice vector and an 8-bit weight vector. It contains four $2\text{-bit} \times 8\text{-bit}$ multipliers followed by three adders for accumulation. The activation slice input can be set to zero under the control of a valid signal, allowing blank values to be skipped. A 1-bit shifter is inserted in the partial sum datapath to enable PoT-based requantization.

\begin{figure}[]
    \centering
    \includegraphics[width=\columnwidth]{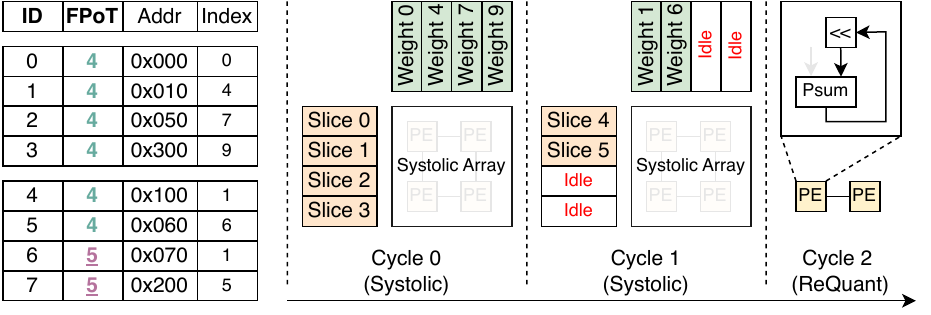} 
    \caption{Illustration of the execution of matrix multiplication and accumulation (MMA) on MPU.}
    \label{fig:mma}
\end{figure}

Fig. \ref{fig:mma} illustrates an example of how matrix multiplication and accumulation (MMA) are executed on MPU. 
A sample block of meta-information for quantized differential activations is shown on the left. In each cycle, the MPU reads four meta-items and then fetches corresponding activation slices and weights according to the \texttt{Addr} and \texttt{Index} fields, respectively. Each activation slice contains $64\times$2-bit differential activations. The control unit in MPC then checks whether all FPoT items are identical. If so, the data is directly streamed into the systolic array, as illustrated in Cycle~0 of Fig. \ref{fig:mma}. In contrast, if not identical, for instance, in Cycle~1 where slice~5 and slice~6 have different FPoT values (4 and 5), the last two lanes of PEs are set to idle. Meanwhile, in the following cycle (Cycle~2), the systolic array stalls 
to perform re-quantization.

As mentioned earlier, our quantization and computation is performed tile by tile.
When accumulating partial sums across tiles, inverse quantization is required. To support this, an inverse quantization unit and an accumulation unit are placed after the PE array. The former is implemented as a vector–scalar multiplication unit, while the latter is implemented as a vector–vector addition unit.

\begin{figure}[]
    \centering
    \includegraphics[width=\columnwidth]{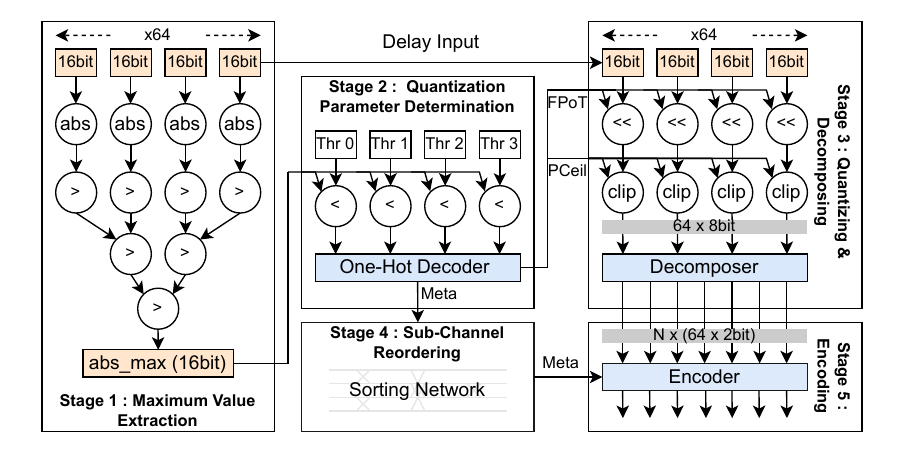} 
    \caption{Design of QPU.}
    \label{fig:qpu}
\end{figure}

\textbf{Design of QPU}.  
We integrate a QPU to support our quantization algorithm as illustrated in Fig. \ref{fig:computation_flow}. The microarchitecture is shown in Fig. \ref{fig:qpu}, consisting of a 5-stage pipeline.  

\textit{Stage 1: Maximum Value Extraction. } 
This stage accepts a 64$\times$16-bit floating-point vector and computes the maximum absolute value. The implementation drops the sign bit
and compares the remaining part using a comparator tree.  

\textit{Stage 2: Quantization Parameter Determination.  }
The output value from Stage 1 is used to determine quantization parameters, which in our algorithm include the FPoT and the corresponding precision ceiling. This is implemented by a bin operation, which classifies each value into a specific range. As described in Sec. \ref{sec:alg-quant}, the quantization parameter directly depends on the range into which the value falls. To determine this, values from Stage 1 are compared in parallel against predefined range boundaries, denoted as $\text{Thr}_0, \text{Thr}_1,\ldots$. The result is encoded as a one-hot vector, which is then decoded to select the correct quantization parameter.  


\textit{Stage 3 and Stage 4: Quantization, Decomposition, and Sub-Channel Reordering.}
These two stages operate in parallel. Once the quantization parameters are determined, Stage 3 quantizes the floating-point inputs into 8-bit values using shifters and clippers. The quantized outputs are then passed to a decomposer, implemented with simple combinational wiring, which reorganizes the 64$\times$8-bit data into 2-bit slices (1, 2, 3, or 4 slices for 2-, 4-, 6-, and 8-bit precision, respectively). 
Meanwhile, Stage 4 sorts the associated meta-information by FPoT using a bitonic sorting network~\cite{BITONIC}.

 \textit{Stage 5: Encoding.} 
The encoder accepts the sorted meta-information from Stage~4, computes the number of shifts, and packs them with corresponding indices.
The resulting encoded data are then combined with the decomposed slices and written consecutively to SRAM or external memory.


\textbf{Design of VPU}. We integrate a VPU to support three types of vector operations. The \emph{element-wise unit} performs vectorized exponential and GELU operations using 64 parallel nonlinear units, implemented with piecewise cubic polynomials and lookup tables. The \emph{reduction unit} adopts a tree structure to accumulate all elements within a vector. The \emph{vector-scalar product unit} multiplies a single scalar with all elements of a vector. All these units are organized into a configurable pipeline, enabling flexible operation support for functions such as GELU and Softmax.

\subsection{Operator Mapping}
To bridge the gap between the operators in DiT models and the micro-operations of our architecture, we describe how the main operators are mapped to our architecture.

\textbf{Linear Operation.}  
The first primary operator is the linear layer, essentially a matrix multiplication between an activation tensor of shape $[m,k]$ and a weight tensor of shape $[k,n]$. We tile the $m$, $n$, and $k$ dimensions with factors of $256$, $512$, and $128$, respectively. These factors are chosen based on the on-chip memory configuration and computation capacity, and they align well with the tensor shapes commonly found in DiTs. In each iteration, activation and weight blocks of sizes $256 \times 128$ and $512 \times 128$ are loaded into the scratchpad memory, then further partitioned into tiles with a size of $64 \times 128$ and distributed across cores for MMA. 
Moreover, for differential activations, the reverse differential operation is fused by preloading cached activations into SRAM and adding them to the partial sum derived from the first tile of the matrix multiplication. The GELU activation is also fused with the support of VPU.

\begin{figure}
    \centering
    \includegraphics[width=\columnwidth]{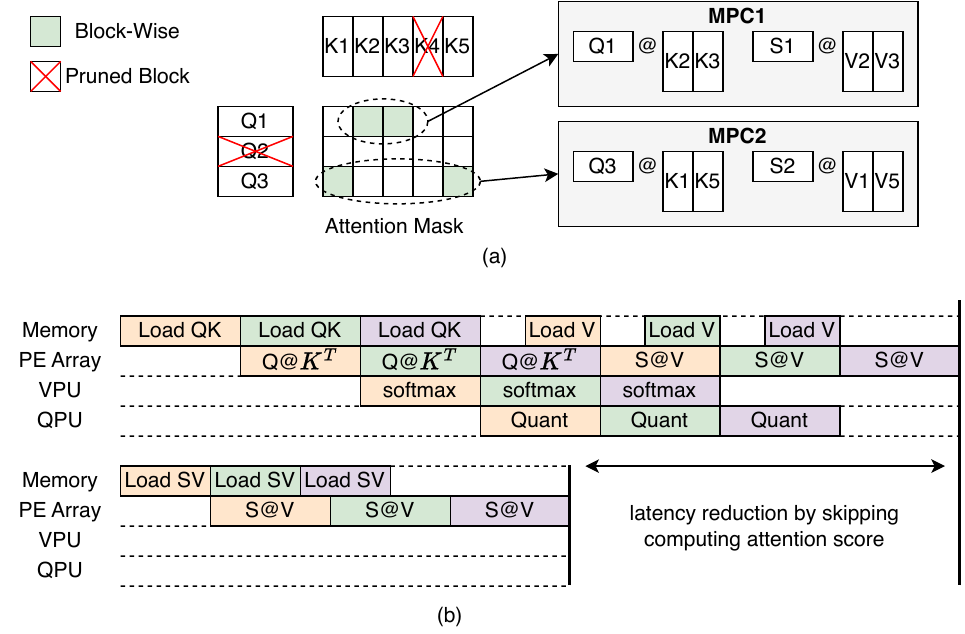} 
    \caption{Illustration of (a) the execution of sparse attention operation and (b) the corresponding pipeline schedule. \(Q_i\), \(K_i\) and \(V_i\) represent query, key, and value vectors, \(S_i\) represents the softmax result.} 
    \label{fig:attn_schedule}
\end{figure}


\textbf{Attention Operation.}  
For the attention operator with SAR, DSTAR employs a static scheduling strategy guided by the attention mask. The key idea is to prune unused $Q$ and $K$ blocks and distribute the remaining blocks across all MPCs. As shown in Fig.~\ref{fig:attn_schedule}~(a), in this block-wise attention mask, blocks $Q_2$ and $K_4$ are never accessed and can therefore be eliminated. The remaining blocks that share the same $Q$ block are assigned to the same MPC.

Within each MPC, we adopt a dataflow similar to FlashAttention~\cite{FLASHATTENTION}. As illustrated in the top part of Fig.~\ref{fig:attn_schedule}~(b), the computation of attention proceeds as follows: load a $QK$ tile, perform matrix multiplication, apply softmax and quantization, then load the $V$ tile, followed by another MMA 
Note that all MMA operations in attention are performed under INT8 precision. We perform quantization on activations using FMDQ, but with INT8 precision; these 8-bit values are then decomposed into bit slices as discussed in Section~\ref{sec:alg-quant}.  
Because the PE array, VPU, and QPU are designed as separate units, the result of $QK^T$ must be transferred to VPU for softmax computation and to QPU for quantization, which introduces idle time in the PE array. To reduce this overhead, multiple $Q$ and $K$ blocks are scheduled into the PE array during these idle periods, as shown in the blocks with different colors in Fig.~\ref{fig:attn_schedule}~(b). During the reuse phase in SAR, since attention scores are reused, neither $QK^T$ computation nor softmax is required. The scheduling switches to the pipeline illustrated in the bottom part of Fig.~\ref{fig:attn_schedule}~(b).






\section{Evaluation}
\label{sec:evaluation}

\begin{table*}[]
  \centering
  
  \caption{Evaluation of model performance.}

\begin{threeparttable}
  
  \resizebox{\textwidth}{!}{
    \begin{tabular}{p{5em}cccccccccccccccccc}
    \toprule
    \multicolumn{1}{c}{\textbf{Model}} & \multicolumn{2}{c}{\textbf{DiT-XL}} & \multicolumn{3}{c}{\textbf{PixArt-Sigma}} & \multicolumn{3}{c}{\textbf{SD3.5-Medium}} & \multicolumn{3}{c}{\textbf{SD3.5-Large}} & \multicolumn{3}{c}{\textbf{Flux.1-Dev}} & \multicolumn{3}{c}{\textbf{Flux.1-Schnell}} & \multicolumn{1}{c}{\textbf{Latte}} \\

     \multicolumn{1}{c}{\textbf{Parameter Size}} & \multicolumn{2}{c}{0.675B} & \multicolumn{3}{c}{0.6B} & \multicolumn{3}{c}{2.5B} & \multicolumn{3}{c}{8.1B} & \multicolumn{3}{c}{12B} & \multicolumn{3}{c}{12B} & \multicolumn{1}{c}{0.675B} \\
    
    \multicolumn{1}{c}{\textbf{Task}} & \multicolumn{2}{c}{Class to Image} & \multicolumn{3}{c}{Text to Image} & \multicolumn{3}{c}{Text to Image} & \multicolumn{3}{c}{Text to Image} & \multicolumn{3}{c}{Text to Image} & \multicolumn{3}{c}{Text to Image} & \multicolumn{1}{c}{Text to Video} \\
    \multicolumn{1}{c}{\textbf{Iterations}} & \multicolumn{2}{c}{50} & \multicolumn{3}{c}{28} & \multicolumn{3}{c}{28} & \multicolumn{3}{c}{28} & \multicolumn{3}{c}{28} & \multicolumn{3}{c}{4} & \multicolumn{1}{c}{50} \\
    \multicolumn{1}{c}{\textbf{Dataset}} & \multicolumn{2}{c}{COCO} & \multicolumn{3}{c}{COCO} & \multicolumn{3}{c}{COCO} & \multicolumn{3}{c}{COCO} & \multicolumn{3}{c}{COCO} & \multicolumn{3}{c}{COCO} & \multicolumn{1}{c}{FETV}  \\
    \midrule
    \multicolumn{18}{c}{\textbf{Parameter for Algorithm Evaluation}} \\
    \midrule
    \multicolumn{1}{c}{SAR} & \multicolumn{2}{c}{Thr=2e-3, Interval=2} & \multicolumn{3}{c}{Thr=1e-4, Interval=2} & \multicolumn{3}{c}{Thre=1e-4, Interval=2} & \multicolumn{3}{c}{Thr=1e-4, Interval=2} & \multicolumn{3}{c}{Thr=1e-4, Interval=2} & \multicolumn{3}{c}{Thr=1e-4, Interval=2} & \multicolumn{1}{c}{\makecell{Thr=1e-4\\ Interval=2}} \\
    \multicolumn{1}{c}{FMDQ} & \multicolumn{2}{c}{ScaleBase=64/256} & \multicolumn{3}{c}{ScaleBase=64/256} & \multicolumn{3}{c}{ScaleBase=64/256} & \multicolumn{3}{c}{ScaleBase=64/256} & \multicolumn{3}{c}{ScaleBase=64/256} & \multicolumn{3}{c}{ScaleBase=64} & \multicolumn{1}{c}{\makecell{ScaleBase=\\64/256}} \\
    \midrule
    \multicolumn{18}{c}{\textbf{Model Performance}} \\
    \midrule
    \multicolumn{1}{c}{} & FID $\downarrow$  & IS $\uparrow$   & FID $\downarrow$  & IS $\uparrow$    & CS $\uparrow$   & FID $\downarrow$  & IS $\uparrow$   & CS  $\uparrow$  & FID $\downarrow$  & IS  $\uparrow$  & CS $\uparrow$   & FID $\downarrow$  & IS $\uparrow$   & CS  $\uparrow$  & FID $\downarrow$  & IS $\uparrow$   & CS  $\uparrow$ & IS  $\uparrow$  \\
\cmidrule{2-19}    \multicolumn{1}{c}{Vanilla} & 202.46 & 41.11 & 103.99 & \textbf{13.72} & 0.313 & 94.10 & \textbf{14.70} & 0.323 & 88.40 & 14.81 & 0.319 & 98.24 & 14.27 & 0.311 & \textbf{96.40} & 14.37 & 0.317 & 26.18 \\

\multicolumn{1}{c}{DITTO} & 200.33
 & 38.19 & 100.84 & 13.23 & 0.314 & 89.88 & 13.60 & 0.326 & 86.76 & 14.46 & 0.323 & 99.33 & 13.39 & 0.313 & 96.92
& 13.65 & 0.317 & 17.96 \\

\multicolumn{1}{c}{Exion} & 200.16 & 35.35 & 107.90 & 10.29 & 0.309 & 105.74 & 11.06 & 0.316 & 124.93 & 10.66 & 0.315 & 95.11 & 12.64 & 0.312 & 90.88 & 13.89 & 0.319 & 18.26 \\

    \multicolumn{1}{c}{DS-SAR} & \textbf{196.39} & 41.77 & 103.93 & 13.48 & 0.312 & 93.30 & 14.20 & 0.322 & 90.62 & 13.82 & 0.319 & 100.91 & 13.89 & 0.312 & 98.37 & 14.70 & \textbf{0.320} & 26.53 \\
    
    \multicolumn{1}{c}{DS-FMDQ} & 204.89 & 42.73 & 101.78 & 13.36 & \textbf{0.316} & 91.16 & 14.64 & \textbf{0.324} & \textbf{86.70} & 15.12 & \textbf{0.321} & \textbf{97.74} & \textbf{14.44} & 0.311 & 96.48 & 14.36 & 0.317 & 26.54 \\
    \multicolumn{1}{c}{DS-FMDQ-DYN} & 204.55 & \textbf{42.89} & 101.31 & \textbf{13.72} & \textbf{0.316} & 90.64 & 14.53 & \textbf{0.324} & 87.58 & \textbf{15.26} & \textbf{0.321} & 99.63 & 13.89 & 0.312 & 97.17 & 14.37 & 0.316 & 26.35 \\
    \multicolumn{1}{c}{DS-All}   & 198.64 & 41.96 & \textbf{100.61} & 13.63 & \textbf{0.316} & \textbf{90.15} & 14.16 & \textbf{0.324} & 88.43 & 14.20 & \textbf{0.321} & 99.73 & 13.85 & \textbf{0.313} & 98.38 & \textbf{14.73 }& 0.319 & \textbf{26.79} \\
    \bottomrule
    \end{tabular}%
    
    }

    \begin{tablenotes}
    \footnotesize
    \item \textbf{FID}: Fréchet Inception Distance; \textbf{IS}: Inception Score; \textbf{CLIP}: CLIP Score measuring image-text alignment.
    \end{tablenotes}

\end{threeparttable}

\label{table:model_perf}

\end{table*}%


\subsection{Experimental Setup}
\label{sec:setup}

\textbf{Algorithm and Workloads.}  
To evaluate the algorithmic performance of DSTAR, we implement it on an NVIDIA A100 GPU \cite{A100}. We select seven DiT-based models covering class-guided and text-guided image generation, as well as text-guided video generation. Specifically, DiT-XL-512 \cite{DITXL} is used for class-guided image generation, and Latte \cite{LATTE} is used for text-to-video generation.
For the text-to-image task, we select models from the three most popular model families, namely PixArt \cite{PIXART}, Stable Diffusion \cite{SD3}, and Flux \cite{FLUX}. 
We evaluate model performance on the COCO dataset \cite{COCO} and the FETV dataset \cite{FETV} for image generation and video generation, respectively, using 1K samples. 
We report Inception Score (IS), Fréchet Inception Distance (FID), and CLIP Score (CS) for each model using open-source evaluation tools \cite{FIDEVALUATE, T2IBENCHMARK}.

\textbf{Architecture.}  
To evaluate the latency of DSTAR, we develop a discrete event simulator based on SimPy \cite{SIMPY}, integrated with Ramulator \cite{RAMULATOR} to assess the performance of our accelerator using memory access traces collected through workload profiling.
For the hardware configuration, we adopt a setup comparable to the NVIDIA A100 GPU. Specifically, 
we adopt 32 MPCs that provide a total of 262 A8W8 TOPS of computational capacity. This is approximately equivalent to 312 TOPS, which is the theoretical FP16 TOPS of the NVIDIA A100 GPU~\cite{A100}.
We incorporate HBM2 with 16 channels, offering a bandwidth of approximately 1500~GB/s.

To evaluate the power and area of our architecture, we first implement the main component of DSTAR in SystemVerilog and synthesize it using Synopsys Design Compiler under the FreePDK 45nm \cite{FREEPDK} technology node at a 1 GHz clock frequency.
Furthermore, we estimate the power and area of the on-chip SRAM using CACTI \cite{CACTI}, which is widely used for estimating on-chip SRAM, such as caches. DRAM power consumption is estimated based on supplier-provided power data. We combine all these power data into our simulator, estimating power consumption over the workloads as mentioned earlier. Additionally, to compare energy efficiency between our design, GPU, and state-of-the-art (SOTA) accelerators, we use DeepScaleTool \cite{DEEPSCALE} to scale the power data of all the accelerators to the same technology node of A100 GPU.  

\textbf{Baselines.}
Our GPU baseline is evaluated on an NVIDIA A100 GPU \cite{A100}. 
We use APIs from Diffusers \cite{DIFFUSERS} for all models.
We insert \texttt{torch.cuda.synchronize()} to record the execution time and use the NVIDIA Nsight tool for power estimation.

We compare DSTAR against four accelerator baselines: S-DMA~\cite{SDMA}, Cambricon-D~\cite{CAMBRICOND}, DITTO~\cite{DITTO}, and EXION~\cite{EXION}. 
To enable a fair comparison, we implement the token merging mechanism used in S-DMA, the outlier-aware schemes adopted in Cambricon-D and DITTO, as well as the FFN-reuse algorithm and Top-$K$ attention pruning used in EXION on a GPU to extract execution traces. 
The simulator reproduces the execution behavior of each accelerator according to the collected traces.
In addition, we include an INT8 systolic array architecture~\cite{TPU}, referred to as INT8-SA. We enhance INT8-SA with group quantization, following the approach used in~\cite{QDIT}, by modifying the PEs to support requantization.
All models are configured to meet a common accuracy standard: less than 5\% degradation in FID/FVD and less than a 1-point drop in IS.
The hardware configurations of all three baselines are scaled to match the TOPS of the DSTAR configuration.

\subsection{Algorithm Evaluation} \label{sec:eval-alg}

Both algorithmic optimizations, namely FMDQ and SAR,
introduce approximation while reducing computational burden.
Therefore, we evaluate their effectiveness in reducing bit-width and computation count, and assess their impact on model accuracy.

\textbf{Model accuracy.} Table~\ref{table:model_perf} shows the impact of our optimization on model performance.
We use the 16-bit floating-point model (\textit{Vanilla}) as the baseline. DSTAR is abbreviated as DS. Starting from \textit{Vanilla}, we progressively apply FMDQ to quantize the differentials between FFN activations in adjacent timesteps and apply SAR to optimize attention operations. 
To evaluate the impact of using a static base in FMDQ, we further introduce a reference variant, denoted as DS-FMDQ-DYN, which adopts a dynamic base for quantization. 
The parameter settings of SAR and FMDQ are summarized in the table. For SAR, $Thr$ denotes the percentage of attention scores to be pruned, and $Interval$ specifies the number of timesteps between recomputation of the attention scores. For FMDQ, $ScaleBase$ represents the minimum quantization scale allowed for each group. In certain models, we assign larger $ScaleBase$ values to the first few layers, as they are more sensitive to quantization error.

\begin{figure}[]
    \centering
    \includegraphics[width=\columnwidth]{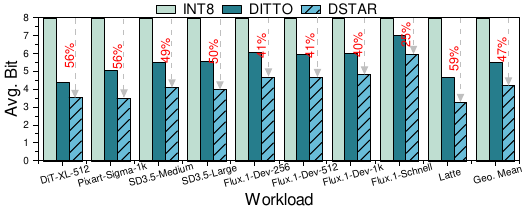} 
    \caption{Comparison of average bit-width. }
    \label{fig:avg_bit}
\end{figure}

\begin{figure}[]
    \centering
    \includegraphics[width=\columnwidth]{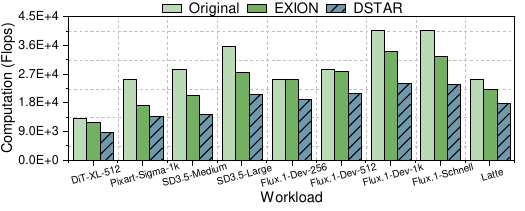} 
    \caption{Comparison of computation reduction.} 
    \label{fig:attn_computation}
\end{figure}

\begin{figure*}[]
    \centering
    \includegraphics[width=0.99\textwidth]{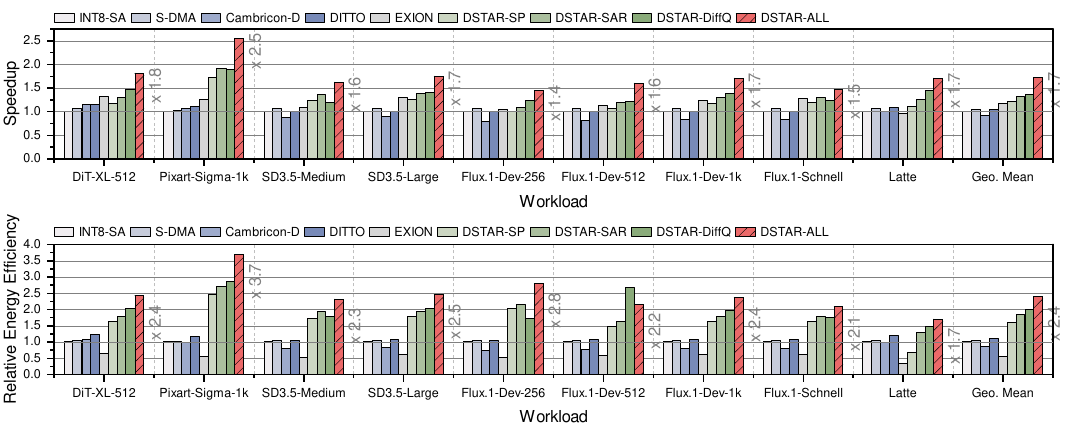} 
    \caption{
    Evaluation of latency speedup and energy efficiency compared with other accelerators. DSTAR-SP, DSTAR-SAR and DSTAR-FMDQ
    refer to DSTAR equipped with sparse attention, sparse attention reuse, and quantization, respectively. DSTAR-ALL enables all optimizations.
    }
    \label{fig:performance}
\end{figure*}

\begin{figure*}[]
    \centering
    \includegraphics[width=0.99\textwidth]{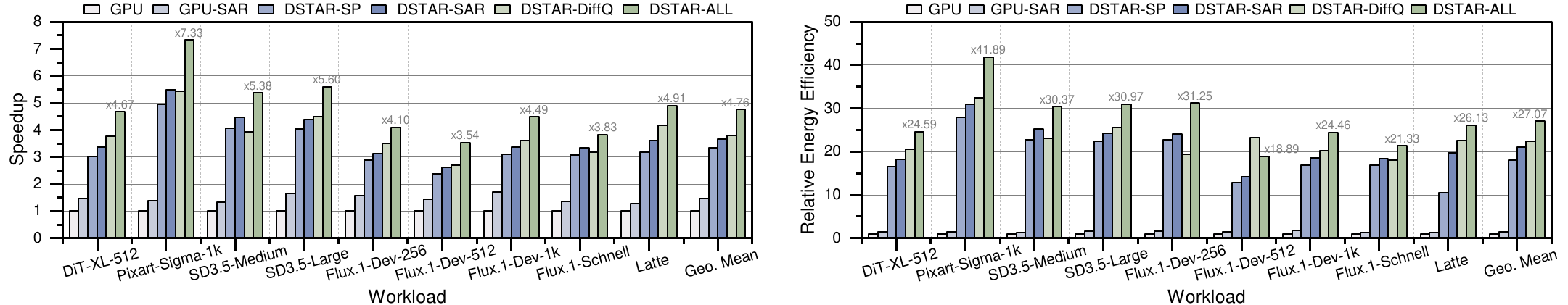} 
    \caption{
    Evaluation of latency speedup and energy efficiency compared with GPU. DSTAR-SP, DSTAR-SAR, DSTAR-FMDQ and DSTAR-ALL are the same as Fig. \ref{fig:performance}.
    }
    \label{fig:performance_gpu}
\end{figure*}

Our optimizations result in negligible performance degradation for most models and, in some cases, even outperform the vanilla model. Although certain models, such as PixArt-Sigma, exhibit relatively larger accuracy drops, the degradation remains within a reasonable range (less than 5\%), as also reported in \cite{TOKENCACHE}.
Furthermore, Figure 19 presents representative generated images from all evaluated models. The visual results show that our optimized approach consistently preserves visual fidelity and is capable of generating high-quality images across all models.




\begin{figure}[]
    \centering
    \includegraphics[width=\columnwidth]{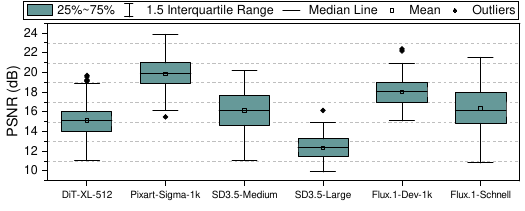} 
    \caption{Evaluation of accuracy sensitivity.}
    \label{fig:sensitivity}
\end{figure}

\begin{figure*}[]
    \centering
    \includegraphics[width=\textwidth]{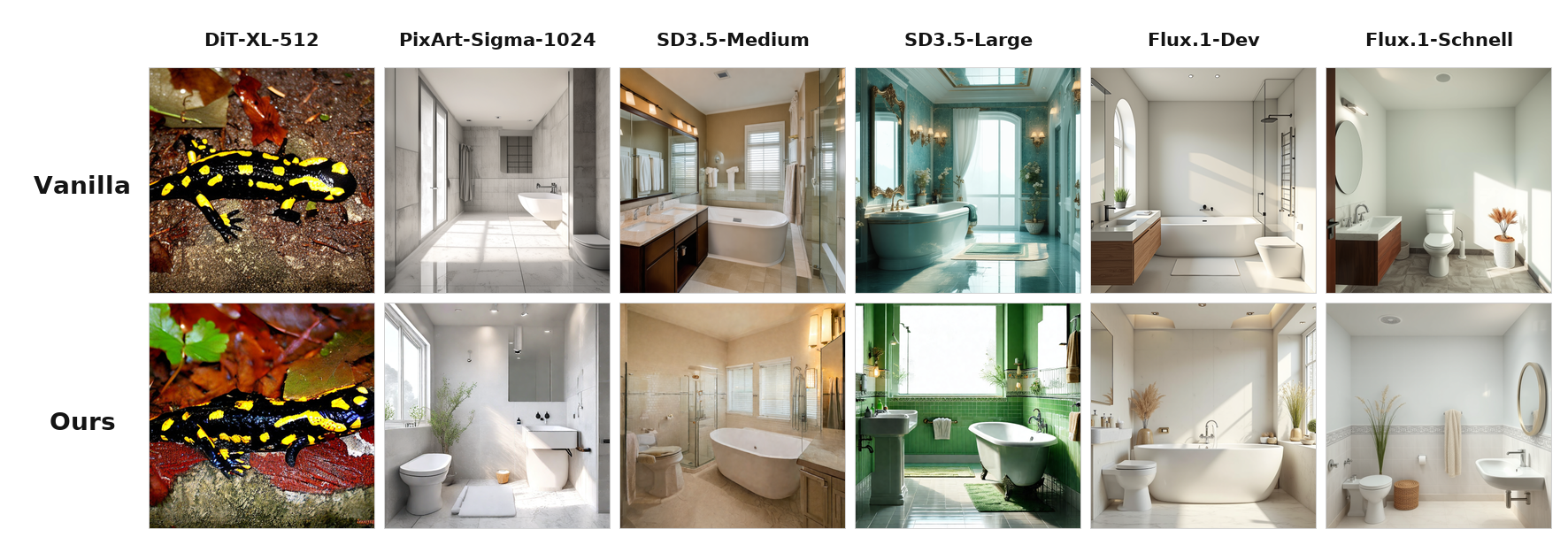} 
    \caption{Visual results generated by the evaluated models.}
    \label{fig:demo}
\end{figure*}

\textbf{Sensitivity Analysis.}
Both SAR and FMDQ rely on static profiling. To evaluate their generalization capability, we conduct a sensitivity study. Specifically, profiling is performed on a training set, while the proposed methods are evaluated on a separate test set using multiple random seeds. Figure~\ref{fig:sensitivity} presents the resulting PSNR values.
The results show that different models exhibit varying levels of sensitivity to static profiling. Models such as PixArt-Sigma-1K  demonstrate limited sensitivity, achieving relatively high PSNR values with low variance across seeds. In contrast, models such as SD3.5-Large achieve lower PSNR values, indicating larger pixel-level deviations. 
Nevertheless, the qualitative results shown in Figure~\ref{fig:demo} indicate that these models still maintain strong visual fidelity and semantic alignment with the input prompts

\textbf{Average Bit-width.}
Figure \ref{fig:avg_bit} presents the average bit-width achieved by FMDQ, compared with the quantization scheme adopted in the state-of-the-art accelerator DITTO \cite{DITTO}. Leveraging fine-grained group selection and mixed precision, FMDQ quantizes differential activations to an average of 4.3 bits. For models with high temporal redundancy, such as Latte, FMDQ further reduces the average bit-width to 3.2 bits—only 41\% of the INT8 baseline. In contrast, DITTO achieves an average bit-width of 5.4 bits. These results highlight FMDQ’s strong capability in exploiting temporal redundancy to reduce quantization precision.

\textbf{Computation reduction of attention layer.} 
Figure \ref{fig:attn_computation} presents the computation reduction of the attention layer when applying Top-K attention 
pruning from EXION \cite{EXION} and the proposed SAR method in DSTAR. SAR achieves up to a 50\% reduction in attention-layer computation, whereas EXION yields at most 32\%. Moreover, in attention-light models such as Flux.1-dev-256, EXION prunes almost no attention scores. This is because SAR’s block-wise pruning pattern is better aligned with the inherent attention structure in DiT, and its reuse mechanism further reduces computation
by skipping redundant operations within the attention layer. Overall, these results highlight the superior effectiveness of SAR in reducing attention-layer computation.

\subsection{Architecture Evaluation}
\label{sec:arch_eval}

\textbf{Latency Speedup.} 
We first compare DSTAR's latency speedup against GPU implementations. We use the implementation in diffusers \cite{DIFFUSERS} library as GPU baseline and additionally implement an SAR kernel using Triton~\cite{TRITON}, denoted as GPU and GPU-SAR, respectively. As shown in Fig.~\ref{fig:performance_gpu}, DSTAR achieves speedups over the GPU baseline ranging from $\times 3.54$ to $\times 7.33$, demonstrating the latency advantage of algorithm--architecture co-design. 
Specifically, GPU-SAR achieves an average $\times 1.46$ speedup over the naive GPU implementation, while DSTAR-SAR achieves an average $\times 3.67$ speedup, highlighting the benefits of the specialized architecture.




We further compare DSTAR with five accelerator baselines described in Section~\ref{sec:setup}, namely INT8-SA, S-DMA, Cambricon-D, DITTO, and EXION. 
As shown in Fig.~\ref{fig:performance}, DSTAR achieves an average $\times$1.63 speedup over S-DMA. Originally designed for U-Net–based diffusion models, S-DMA incorporates a local token-merging strategy. However, when applied to DiT, this strategy introduces large errors because DiT lacks the strong spatial locality found in CNN-based models. Consequently, S-DMA provides only limited performance improvement when transferred to DiT.

Moreover, DSTAR outperforms the state-of-the-art accelerators, Cambricon-D and DITTO, 
by an average speedup of 1.90$\times$ and 1.66$\times$, respectively. Cambricon-D is designed primarily for UNet-based models, and its architecture suffers from PE imbalance.
Optimization in DITTO primarily targets linear layers, leading to limited computation reduction in attention.
Furthermore, DSTAR with only FMDQ achieves 1.50× and 1.32× speedup over Cambricon-D and DITTO, respectively. Both  Cambricon-D and DITTO incorporate coarse-grained quantization mechanisms. As discussed in Section~\ref{sec:temporal-linear}, this introduces great quantization error on large-scale DiTs, making them miss the opportunity to quantize a large portion of values to low bits, even though many are well-suited for lower-precision representation. In contrast, FMDQ employs a finer-grained quantization strategy by dividing the tensor into multiple tiles along the token dimension and applying channel-wise mixed-precision quantization within each tile. This approach increases the proportion of values quantized to low bits, thereby enhancing computational efficiency without compromising accuracy.

Although EXION explores optimization opportunities in both the linear and attention layers, DSTAR still achieves an average 1.48× speedup over it, owing to the combined benefits of FMDQ and SAR, as discussed in Section~\ref{sec:eval-alg}.
FMDQ of DSTAR offers a better accuracy-performance trade-off than EXION which achieves larger average bit-width under the same accuracy.
Moreover, SAR of DSTAR utilizes a pattern better aligned with DiT, and exploits reuse opportunities inherent in DiT,  pruning more computations
compared to EXION while introducing minimal overhead.



\begin{figure}[]
    \centering
    \includegraphics[width=\columnwidth]{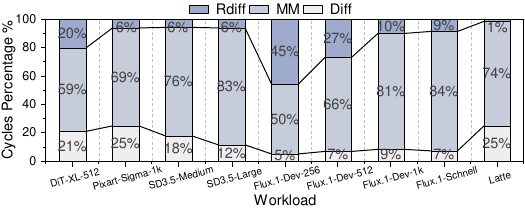} 
    \caption{Breakdown of overhead in FMDQ.}
    \label{fig:fmdq_overhead}
\end{figure}

\begin{figure}[]
    \centering
    \includegraphics[width=\columnwidth]{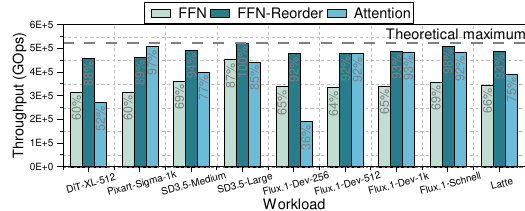} 
    \caption{Throughput and PE utilization.}
    \label{fig:utilization}
\end{figure}

\textbf{Energy Efficiency.}
We also compare the energy efficiency of DSTAR with other hardware designs. 
As shown in Fig.~\ref{fig:performance_gpu}, DSTAR achieves $18.89\times$ to $41.89\times$ higher energy efficiency compared to the A100 GPU. 
This improvement results from both the reduction in 
FLOPs and the power efficiency advantages.
Fig.~\ref{fig:performance} depicts the energy efficiency comparison between DSTAR and prior accelerator designs. DSTAR achieves an average of $2.43\times$ higher energy efficiency than INT8-SA. When comparing with state-of-the-art designs, DSTAR delivers an average of 4.45× higher energy efficiency than EXION, owing to its use of low-precision computation and superior algorithmic optimizations. Moreover, DSTAR achieves 2.87×, 2.27x and 2.21× higher average energy efficiency than DITTO, S-DMA and Cambricon-D, respectively. This improvement primarily stems from the reduced execution time, which significantly lowers the dynamic energy consumption of the PE array,
as well as the reduced number of external memory accesses.


\textbf{Overhead and PE utilization.} Our algorithm introduces additional overhead as well as potential execution imbalance. Thus, we evaluate the impact of these factors. We first analyze the overhead introduced by FMDQ. As shown in Fig.~\ref{fig:fmdq_overhead}, we present the execution cycle breakdown of the three pipeline stages in FMDQ-based matrix multiplication. 
\textit{Diff} denotes the differential quantization stage, \textit{MM} denotes the matrix multiplication stage, and \textit{Rdiff} denotes the reverse differential stage. We observe that, across all evaluated models, \textit{MM} remains the pipeline bottleneck, indicating that the pipelined execution effectively hides most of the additional overhead. 

To evaluate the imbalance and potential PE underutilization introduced by FMDQ and SAR, we further measure the throughput of FFN and attention. The results are shown in Fig.~\ref{fig:utilization}. 
For the naive FFN implementation, the shift operations required by mixed-precision computation and scaling reduce average PE utilization to 67\%. By applying the proposed reorder optimization, this overhead is largely eliminated, enabling all evaluated models to achieve over 88\% of the theoretical peak PE utilization. 

For attention, most models achieve over 75\% PE utilization, except for light-attention models such as DiT-XL-512 and Flux.1-dev-256. Their lower utilization is mainly caused by shorter sequence lengths in attention and the imbalance of loading key/value tensors in sparse attention.

\begin{table}[]
  \centering
  \caption{Area and power consumption breakdown.}
  \resizebox{\columnwidth}{!}{
    \begin{tabular}{ccccc}
    \toprule
    \multicolumn{2}{c}{\textbf{Component}} & \textbf{Parameter} & \textbf{Area (\(mm^2\))} & \textbf{Power (W)} \\
    \midrule
    \multirow{4}[2]{*}{MPC} & MPU & 64$\times$64 PE array & 3.67 & 3.4307 \\
          & VPU   & SIMD64 & 0.05 & 0.0502 \\
          & QPU   & SIMD64 & 0.10 & 0.0966 \\
          & Unified Buffer & 80 KB, 10 Banks & 1.51 & 0.1320 \\
    \midrule
    \multicolumn{2}{c}{Global Scratchpad} & 512 KB, 32 Banks & 6.49 & 0.0457 \\
    \midrule
    \multicolumn{2}{c}{\textbf{Total}} & 32 MPCs & 176.75 & 118.75 \\
    \bottomrule
    \end{tabular}%
    }
  \label{tab:addlabel}%
  \label{tab:ap_breakdown}%
\end{table}%

\textbf{Area and Power Consumption.}
Table \ref{tab:ap_breakdown} presents the area and power breakdown of the DSTAR hardware architecture implemented in 45nm technology node.
With the configuration of 32 MPCs and a 512KB global scratchpad, the DSTAR hardware accelerator occupies a total area of 176.75mm\(^2\) and consumes the power of around 118.75W.
These values are significantly smaller than the 826mm\(^2\) die size and 250W thermal design power (TDP) of the Nvidia A100 GPU \cite{A100AP}, which is fabricated in 7nm technology node.
Among all components, the PE array occupies the majority of the area and power consumption, consuming 68.37\% of area and 92.48\% of power consumption, respectively. 
Regarding the QPU, it accounts for only 1.8\% of the area and 2.6\% of the power consumption of a single MPC, indicating its negligible overhead when processing our quantization algorithm.



\begin{figure}[]
    \centering
    \includegraphics[width=\columnwidth]{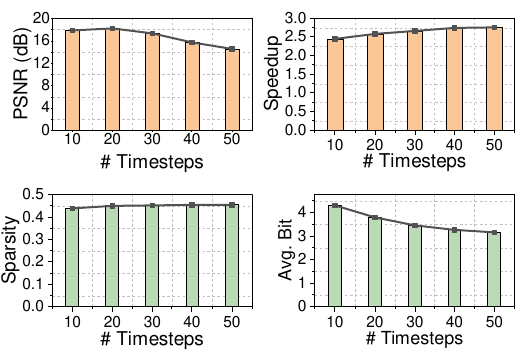} 
    \caption{Ablation study across different timestep settings.}
    \label{fig:timestep_ablation}
\end{figure}



\subsection{Ablation Study of Timestep}

To evaluate the effectiveness of DSTAR under different diffusion sampling schedules, we conduct a timestep ablation study on PixArt-Sigma 
The number of denoising timesteps ranges from 10 to 50. Figure~\ref{fig:timestep_ablation} reports the image quality (PSNR), overall speedup over the GPU baseline, SAR sparsity ratio, and the average quantization bitwidth of FMDQ under different timestep settings.

The results show that DSTAR maintains stable accuracy across a wide range of timesteps. The PSNR varies only from 14.53 dB to 18.24 dB, demonstrating the robustness of our approach under different sampling schedules. Similarly, the speedup remains stable, ranging from 2.45$\times$ to 2.76$\times$, indicating that DSTAR delivers performance benefits regardless of the number of denoising steps.

Interestingly, the speedup slightly increases as the number of timesteps grows. To better understand this trend, we further analyze the SAR sparsity ratio and the average bitwidth of FMDQ. As shown in Figure~\ref{fig:timestep_ablation}, the sparsity ratio of SAR remains nearly constant across different timestep settings. This is expected because the semantic sparsity exploited by SAR is largely independent of the total number of denoising steps. In contrast, the average bitwidth of FMDQ decreases as the number of timesteps increases. As discussed previously, a larger number of denoising steps leads to higher similarity between activations from adjacent timesteps, resulting in smaller activation differences and therefore lower differential quantization precision requirements. 


\section{Conclusion}


In this work, we introduce DSTAR, a software-hardware co-design framework for DiT inference acceleration by effectively exploiting both spatial and temporal redundancies. 
Algorithmic and hardware optimizations are proposed to improve the DiT inference efficiency while maintaining high generation quality of models.

\begin{acks}
This work was supported by the National Natural Science Foundation of China under Grant 62472273 and Grant 62232015.
\end{acks}


\bibliographystyle{ACM-Reference-Format}
\bibliography{refs}

\end{document}